\documentstyle[aps,pre,epsf,twocolumn,eqsecnum]{revtex}

\newcommand{\ie}{i.e.}
\newcommand{\et}{\textit{et.~al.}}

\begin{document}

\wideabs{
\title{Dynamic ductile to brittle transition \\ in a one-dimensional
  model of viscoplasticity}

\author{Alexander E.~Lobkovsky$^\dag$ and J.~S.~Langer$^\ddag$}

\address{$^\dag$Institute for Theoretical Physics, University of
  California, Santa Barbara, CA 93106 \\ $^\ddag$Department of
  Physics, University of California, Santa Barbara, CA 93106}

\draft \date{\today}

\maketitle

\begin{abstract}
  We study two closely related, nonlinear models of a viscoplastic
  solid.  These models capture essential features of plasticity over a
  wide range of strain rates and applied stresses.  They exhibit
  inelastic strain relaxation and steady flow above a well defined
  yield stress.  In this paper, we describe a first step in exploring
  the implications of these models for theories of fracture and
  related phenomena.  We consider a one dimensional problem of
  decohesion from a substrate of a membrane that obeys the
  viscoplastic constitutive equations that we have constructed.  We
  find that, quite generally, when the yield stress becomes smaller
  than some threshold value, the energy required for steady decohesion
  becomes a non-monotonic function of the decohesion speed.  As a
  consequence, steady state decohesion at certain speeds becomes
  unstable.  We believe that these results are relevant to
  understanding the ductile to brittle transition as well as fracture
  stability.
\end{abstract}
}

\section{Introduction}
\label{sec:intro}

A wide range of evidence points toward the necessity of including
plastic deformation near a crack tip among the relevant degrees of
freedom in theories of dynamic brittle fracture.  Our own ideas about
this issue emerge from our recent attempt to study fracture stability
using a class of cohesive-zone models in which such deformations are
necessarily absent.  As described in our report on this project
\cite{monster2}, we discovered both mathematical and physical
difficulties that, so far as we could tell, can be overcome only by
introducing tip blunting or, perhaps, a spatially extended process
zone in which irreversible deformations of the brittle solid are
taking place.

A successful theory of dynamic fracture also must explain failure of
materials that can flow plastically.  Stroh \cite{stroh} understood
that cleavage in such materials is an inherently dynamic process in
which plastic flow is slow enough that stresses can increase to large
values near the crack tip.  In order to explain slower ductile crack
propagation, McClintock \cite{mcclintock} introduced a novel void
nucleation and coalescence mechanism.  At present, theories of these
two failure mechanisms are separate and incompatible with one another.

Attempts to explain the differences between ductile to brittle
fracture usually focus on the emission and mobility of dislocations
near the crack tip \cite{rice_thompson,ashby}.  A different
microscopic mechanism must be responsible for the ductile to brittle
transition in noncrystalline materials such as toughened
thermoplastics \cite{bucknall,vukhahn}.  The study of Freund and
Hutchinson aimed at understanding dynamic fracture within a theory of
continuum plasticity \cite{freund_hutch85,mataga}.  Under the
assumption of elastic strain-rate dominance near the tip, they found a
non-monotonic dependence of the fracture toughness on the crack speed.
They used an idealized viscoplastic constitutive law, however, in
which the plastic strain rate is identically zero below a yield stress
and responds instantaneously to changes in the stress.  The condition
of elastic strain-rate dominance breaks down for slow cracks, and thus
the analysis of Freund and Hutchinson cannot be generally valid.  A
different approach by Freund's \cite{freund_book} has been to use a
rate-dependent cohesive-zone model \cite{glennie} and to consider two
separate fracture criteria, one based on stress and the other on
displacement.  This is the class of models that we found to be
ill-posed for our stability analyses, presumably because they omit
essential features of an extended plastic process zone.

In a recent study, M.~L.~Falk and one of the present authors (JSL)
have proposed a theory of viscoplasticity in amorphous materials
\cite{michael} (denoted FL in what follows) in which the equations of
linear elastodynamics are supplemented by nonlinear equations of
motion for a set of internal state variables.  This theory is based
directly on molecular-dynamics simulations which revealed the
existence of localized weak regions, called ``shear transformation
zones.''  The new internal state variables describe the populations
of these zones, and the nonlinear equations describe how those
populations determine the time-dependent elastic and plastic behavior
of the material.

In the picture presented in FL, the shear transformation zones are
two-state systems, and transitions between those states produce
increments in the plastic strain.  Because a zone that has transformed
once cannot transform again in the same direction, the plastic strain
remains bounded when the applied stress is small.  An additional
assumption of FL is that the zones are created and annihilated at
rates determined by the inelastic shear rate.  It is this process, in
which new zones replace old ones, that produces persistent plastic
flow at sufficiently large stresses.  Plasticity is a fully dynamic
phenomenon in this theory.  It occurs in a well defined way, depending
on the state of the system, in response to time dependent
perturbations.  Thus this new theory may be capable of describing both
brittleness and ductility in fracture.

We report here on our initial attempts to incorporate some of the
basic features of the FL theory into a model of fracture.  As a first
step in exploring the implications of this theory, we examine a
one-dimensional model of decohesion of a thin membrane from a
substrate, where the membrane obeys a simplified version of the FL
viscoplastic constitutive equations.  The membrane is pulled from the
substrate by weak springs, and the point of decohesion propagates at
constant speed, like a crack tip.  As noted by Barber \et\ 
\cite{barber} and Marder and Gross \cite{marder}, the inverse
stiffness of the driving springs in such models is analogous, at least
in some ways, to the width of the strip in two-dimensional mode-I
fracture.  This effective width plays an important role in our
interpretation of the brittle-ductile transition in this system.  The
one-dimensional nature of the model is, of course, a significant
limitation.

This article is organized as follows.  We introduce our simplified FL
model in Section \ref{sec:model}.  In particular, we introduce two
different versions of the nonlinear term that determines the plastic
yield stress, and we examine the time dependence of the system near
the onset of plastic flow for these two cases.  In Section
\ref{sec:linear}, we discuss the linear version of this model and show
that it corresponds to conventional viscoelasticity with inelastic
deformation but no persistent plastic flow.  Then, in Section
\ref{sec:nonlinear}, we describe both analytic and numerical studies
of the nonlinear model, and show how the two rate-dependent mechanisms
introduced in Section \ref{sec:model} produce different versions of a
brittle-ductile transition.  We conclude in Section
\ref{sec:discussion} with a discussion of how the lessons learned from
this simple class of models might be applicable to more realistic
situations.

\section{One-dimensional model of decohesion}
\label{sec:model}

\begin{figure}[htbp]
  \begin{center}
    \centerline{\epsfxsize=3.2in \epsfbox{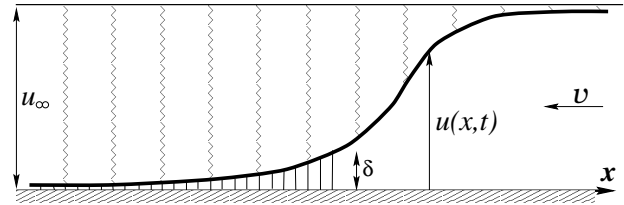}}
    \vspace{0.1in}
    \caption{A model of one dimensional decohesion driven by weak
      springs.}
    \label{fig:config}
  \end{center}
\end{figure}

We consider a thin membrane decohering from a substrate as shown in
Fig.~\ref{fig:config}.  The system has translational symmetry in the
direction perpendicular to the plane of the Figure.  As an additional
simplifying approximation, we allow the membrane to move only in the
direction perpendicular to the substrate.  The configuration of the
membrane is thus described by its displacement $u(x,t)$, a
non-negative function of position $x$ and time $t$.  A cohesive force,
\begin{equation}
  f = \cases{-\kappa^2 u & for $0 <u <\delta$ \cr 0 & otherwise,\cr}
\end{equation}
with a finite range $\delta$, attracts the membrane to the substrate.
Decohesion is driven by weak springs of strength $\alpha^2$ whose
relaxed positions are at $u = u_\infty$.  The total strain in the
membrane is $\epsilon_{tot} = \partial u/\partial x$.  This strain
and, equivalently, the displacement $u$ consist of additive elastic
and plastic parts:
\begin{equation}
  \label{eq:straindef}
  u = u_{el} + u_{pl}; \qquad \epsilon_{tot} = {\partial u_{el} \over
    \partial x} + {\partial u_{pl} \over \partial x}.
\end{equation}
For later notational simplicity, we write
\begin{equation}
  {\partial u_{pl} \over \partial x} = \epsilon.
\end{equation}
By definition, the elastic part of the strain is linearly proportional
to the stress, $\sigma = 2\mu (\partial u_{el}/\partial x)$, where
$\mu$ is the shear modulus.  Then the equation of motion for the
membrane is
\begin{equation}
  \label{eq:motion}
  \rho \ddot u = {\partial \sigma \over \partial x} - \kappa^2 u
  \,\Theta(\delta - u) - \alpha^2 (u - u_\infty),
\end{equation}
where $\rho$ is the linear mass density and $\Theta(\cdot)$ is the
Heaviside step function.  Dots denote time derivatives.

The equation of motion for the plastic strain that we shall use here
is:
\begin{equation}
  \label{eq:plastic_falk}
  \dot \epsilon = {1 \over \tau}\Bigl(\lambda \sigma - \Delta\Bigr).
\end{equation}
This is a simplified version of Eq.~(3.14) in FL. (We have evaluated
the right-hand side of the latter equation in the limit of small
stress $\sigma$ and have set $n_{\Delta} = \Delta$, $n_{tot} =
n_{\infty}$.)  By making this small-$\sigma$ approximation, we lose
some of the memory effects that were obtained in FL via a strongly
nonlinear $\sigma$-dependence of the rate factors in the equation for
$\dot \epsilon$.  We believe that the absence of those effects makes
only quantitative and not qualitative differences  in the results to
be presented here.  However, that point may require further
investigation.

Our single state variable, $\Delta(x,t)$, is a measure of the
imbalance in the populations of the two-state systems.  The first term
on the right-hand side of (\ref{eq:plastic_falk}), $\lambda \sigma/
\tau$, is the usual linear relation between the plastic strain rate
and the stress.  The second term, $-\Delta/\tau$, is the rate at which
these two-state systems transform spontaneously from their ``forward''
to their ``backward'' states, and is therefore a negative contribution
to $\dot \epsilon$.

Our equation of motion for $\Delta$ has the form:
\begin{equation}
  \label{eq:delta_motion}
  \dot\Delta = \dot\epsilon - {\cal F}(\dot \epsilon,\sigma) \Delta.
\end{equation}
This is exactly the same as Eq.~(3.15) in FL except that we have not
yet specified the strain-rate dependent coupling ${\cal F}$ between
$\dot\Delta$ and $\Delta$.  The first term on the right hand side of
(\ref{eq:delta_motion}) simply expresses our assumption that the
transitions within the two-state systems correspond to increments in
plastic strain.  The second, \ie\ the nonlinear term, is the effect of
creation and annihilation of these zones and, as we shall see
immediately, is responsible for the existence of a finite plastic
yield stress.

To illustrate the properties of the nonlinear term in
(\ref{eq:delta_motion}), we consider two plausible forms of ${\cal F}$
that produce qualitatively different dynamic decohesion in certain
regimes.  The form of ${\cal F}$ is restricted by the assumption that
it must vanish when the plastic strain rate vanishes.  It is thus
proportional to some power of $\dot\epsilon$.  (In higher dimensions,
we would also require rotational invariance.)  The first form that we
shall examine is the same as that used in FL: ${\cal F}_1 = \gamma_1
\dot \epsilon \sigma$.  Here, the coupling ${\cal F}_1$ is
proportional to the local rate of plastic energy dissipation.  We call
this Model 1.  Note that ${\cal F}_1$ can be negative in some
circumstances.  Our second possibility, Model 2, is one in which only
the local plastic strain rate controls the evolution of $\Delta$, in
which case the simplest choice is ${\cal F}_2 = \gamma_2 \dot
\epsilon^2$.

\begin{figure}[htbp]
  \begin{center}
    \centerline{\epsfxsize=3.2in \epsfbox{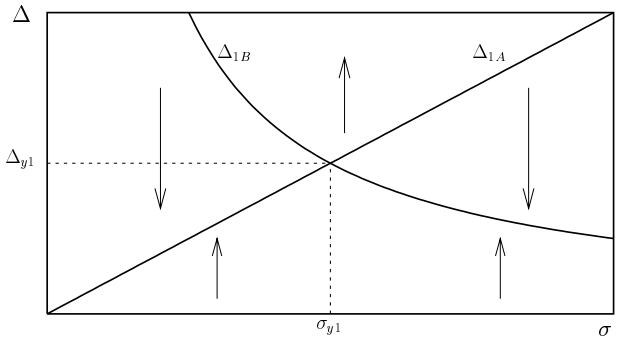}}
    \vspace{0.1in}
    \caption{The steady state values of $\Delta$ in Model 1 as
      functions of $\sigma.$ The two curves $\Delta_{1A} = \lambda
      \sigma$ and $\Delta_{1B} = 1/\gamma_1 \sigma$ cross at
      $(\sigma_{y1}, \Delta_{y1}).$  The arrows indicate the direction
      of motion of $\Delta$ for fixed $\sigma.$}
    \label{fig:mod1_flow}
  \end{center}
\end{figure}
We explore first the behavior of Model 1.  Substituting the expression
for $\dot \epsilon$ from (\ref{eq:plastic_falk}) into
(\ref{eq:delta_motion}), we obtain
\begin{equation}
  \label{eq:falk_Delta_evol}
  \dot \Delta = {1 \over \tau}\,\left(\lambda \sigma - \Delta \right)
  \left(1 - \gamma_1 \sigma \Delta \right).
\end{equation}
For constant stress $\sigma$, there are two stationary solutions:
\begin{equation}
  \label{eq:falk_deltas}
  \Delta = \Delta_{1A} = \lambda\sigma, ~~~\text{and}~~~\Delta =
  \Delta_{1B} = {1 \over \gamma_1\sigma},
\end{equation}
which, as shown in Figure \ref{fig:mod1_flow}, cross at
\begin{equation}
  \label{eq:falk_critical}
  \sigma = \sigma_{y1} = {1 \over \sqrt{\gamma_1\lambda}},~~~~\Delta =
  \Delta_{y1} = \sqrt{\lambda \over \gamma_1}.
\end{equation}
At any fixed $\sigma$, only one of these stationary solutions is
stable against perturbations. For $\sigma < \sigma_{y1}$, the
stationary solution $\Delta_{1A}$ with $\dot\epsilon = 0$ is stable.
For $\sigma >\sigma_{y1}$, on the other hand, the stable stationary
solution $\Delta = \Delta_{1B}$ is a flowing steady state with
\begin{equation}
  \dot\epsilon = {\lambda \over \sigma \tau} (\sigma^2 -
  \sigma_{y1}^2).
\end{equation}
This rate vanishes when the stress approaches yield from above.
However, the relaxation time $\tau_1$ for perturbations away from the
flowing state diverges for $\sigma$ near $\sigma_{y1}$:
\begin{equation}
  \label{eq:falk_tau1}
  \tau_1 = {\sigma_{y1}^2 \tau \over |\sigma^2 - \sigma_{y1}^2|}.
\end{equation}
This is quite unlike the conventional elastic-ideally plastic solid in
which this relaxation time is zero by definition.

A possibly unphysical feature of this model is that, for some initial
conditions, $\Delta$ may increase indefinitely as a function of time.
It is easy to see from (\ref{eq:falk_Delta_evol}), however, that if
$\Delta < \Delta_{y1}$ at any time, it will remain so for all other
times regardless of the stress history.

\begin{figure}[htbp]
  \begin{center}
    \centerline{\epsfxsize=3.2in \epsfbox{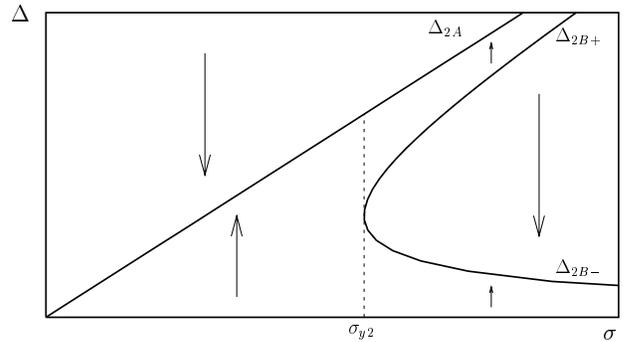}}
    \vspace{0.1in}
    \caption{Steady state values of $\Delta$ as functions of $\sigma$
      for Model 2.  One of the three, $\Delta_{2B+},$ is unstable to
      small perturbations.  Since the steady state curves never cross,
      the system remains close to the non-flowing steady state
      $\Delta_{2A}$ if the stress increases slowly enough.  The
      meaning of the arrows is the same as in
      Fig.~\ref{fig:mod1_flow}.}
    \label{fig:mod2_flow}
  \end{center}
\end{figure}
Model 2 exhibits an important qualitative difference in its behavior.
Let us perform the analysis of the preceding paragraphs using ${\cal
  F}_2$. Substituting (\ref{eq:plastic_falk}) into
(\ref{eq:delta_motion}), we obtain
\begin{equation}
  \dot \Delta = {1 \over \tau}\,\left(\lambda \sigma - \Delta \right)
  \left({\gamma_2\over \tau}\Delta^2 - {\gamma_2\lambda
      \sigma\over\tau} \Delta + 1\right).
\end{equation}
We again look for the stationary states $\dot \Delta = 0$.  The
situation is shown in Fig.~\ref{fig:mod2_flow}.  In this case, the
state with $\Delta = \Delta_{2A} = \lambda \sigma$ and no plastic
flow, $\dot \epsilon = 0$, is stable for all $\sigma$.  As seen in
Fig.~\ref{fig:mod2_flow}, it never intersects any other state in the
$\Delta$-$\sigma$ plane.  For
\begin{equation}
  \label{eq:bad_critical}
  \sigma > \sigma_{y2} = {2 \over \lambda} \sqrt{\tau \over \gamma_2},
\end{equation}
two new stationary states appear with
\begin{equation}
  \label{eq:Delta_overcrit}
  \Delta = \Delta_{2B\pm} =  {\lambda \over 2}\left(\sigma \pm
    \sqrt{\sigma^2 - \sigma_{y2}^2}\right).
\end{equation}
The state with $\Delta = \Delta_{2B-}$ is the stable one of the pair.
The plastic strain rate in this stationary state is non-zero:
\begin{equation}
  \dot \epsilon = {\lambda\over 2\tau} \left(\sigma + \sqrt{\sigma^2 -
      \sigma_{y2}^2}\right).~~~~(\textrm{steady state,}~~~\sigma >
  \sigma_{y2}).
\end{equation}  
The characteristic decay time $\tau_2$ of perturbations around the
flowing stationary state again diverges at the yield stress, although
the divergence is not as strong as in Model 1:
\begin{equation}
  \label{eq:decay_flow}
  \tau_2 = {\tau \over 2} {\sigma_{y2}^2 \over \left(\sigma +
      \sqrt{\sigma^2 - \sigma_{y2}^2}\right)\sqrt{\sigma^2 -
      \sigma_{y2}^2}},~~~~(\sigma > \sigma_{y2}).
\end{equation}

The two nonlinear models exhibit a number of similar features.  Most
importantly, steady plastic flow in response to a stress above a yield
level is a natural consequence of the dynamical constitutive
equations.  The flow has a non-zero response time to changes in the
stress. It can also be shown that inelastic strain is partially
recovered in both models. However, there are several important
differences between the two nonlinear models. First, the steady flow
rate does not vanish in Model 2 at $\sigma = \sigma_{y2}$.  Second,
``runaway'' behavior cannot occur in that model since $\dot\Delta < 0$
for $\Delta > \Delta_{2A}$. And finally, for stresses greater than the
yield stress, there are two stable stationary solutions in Model 2 as
opposed to only one in Model 1.  Which one of these is selected by the
system depends on the stress history.  For example, only the
non-flowing state in Model 2, with $\Delta = \Delta_{2A}$, occurs if
the stress is increased slowly enough, \ie\ when $\dot \sigma/\sigma
\ll 1/\tau$.  As we shall see, this distinction between the models
leads to qualitatively different behaviors at small speeds.

It is convenient at this point to convert to dimensionless variables
in which all lengths are measured in units of the range of the
cohesive interaction $\delta$, time is measured in units of $\delta
\sqrt{\rho/2\mu}$, and stress in units of $2\mu$.  For simplicity, we
continue to use the symbols $u$ and $\sigma$ for our dimensionless
displacements and stresses.  We also restrict our attention to
steady-state solutions moving in the negative $x$ direction with speed
$v$.  All functions of $x$ and $t$ in the frame of reference moving
with the decohesion front depend only on the combination $x' = x +
vt$.  Without loss of generality, we set $x' = 0$ at the point of
decohesion where, in these units, $u = 1$.  Then, for simplicity, we
set $x' = x$.

Our equations of motion now have the form:
\begin{eqnarray}
  \label{eq:steady_motion}
  v^2 (u''_{el} + u''_{pl}) & = & u''_{el} - \kappa^2 (u_{el} +
  u_{pl}) \Theta(-x) - \cr
  && \alpha^2 (u_{el} + u_{pl} - u_\infty), \\
  \label{eq:steady_motion2} 
  v u''_{pl} & = & {1 \over \tau} \left(\lambda u'_{el} - \Delta
  \right), \\
  \label{eq:steady_motion3} 
  v\Delta' & = & v u''_{pl} - {\cal F}(v u''_{pl}, u'_{el})\Delta,
\end{eqnarray}
where primes denote differentiation with respect to $x$, and
\begin{eqnarray}
  {\cal F}_1 & = & \gamma_1 v u''_{pl} u'_{el};\\
  {\cal F}_2 & = & \gamma_2 (v u''_{pl})^2.
\end{eqnarray}

Finally, we derive an expression for the decohesion resistance $G(v)$,
which is the work that the driving springs perform on the membrane per
unit length of advance of the decohesion front.  Since the driving
springs relax to their equilibrium length far behind the decohesion
front, all of their stored elastic energy ahead of the front must be
dissipated in the decohesion process.  Thus, the total work done must
be
\begin{equation}
  \label{eq:G}
  G(v) = {1 \over 2} \alpha^2 u_\infty^2.
\end{equation}
If we multiply (\ref{eq:steady_motion}) by $u'(x)$ and integrate over
$x$, we obtain
\begin{equation}
  \label{eq:energy_balance}
  G(v) = {1 \over 2} \,\kappa^2 + \int dx \, u''_{pl} u'_{el} + {1
    \over 2} \left.(u'_{pl})^2\right|_{x = \infty}.
\end{equation}
The first term on the right-hand side of (\ref{eq:energy_balance}) is
clearly the energy spent in breaking cohesive bonds.  The second term
is the energy dissipated in plastic work in the neighborhood of the
decohesion front.  The third term is the energy locked into the
plastic wake left by the decohesion. The decohesion front in this
model may leave a residual plastic strain behind it.  Our problem is
to compute $G(v)$ explicitly as a function of $v$ and then use
(\ref{eq:G}) to determine $v$ as a function of the driving force
$\alpha u_{\infty}$.

\section{Linear analysis}
\label{sec:linear}

Before going ahead with an analysis of this nonlinear model of
viscoplasticity, it will be useful to look briefly at the linear case
in which we set ${\cal F} = 0$ in (\ref{eq:delta_motion}).  Then we
have $\Delta = \epsilon = u'_{pl}$; and the equation of motion for
$u_{pl}$ is
\begin{equation}
  \label{eq:lindotu}
  \dot u_{pl} = v u'_{pl} = {1 \over \tau} (\lambda u_{el} - u_{pl}).
\end{equation}
The remaining equation of motion is (\ref{eq:motion}) or,
equivalently, (\ref{eq:steady_motion}).

Far away from the region where decohesion is taking place, our system
is translationally invariant, and we can compute a dispersion relation
for waves of the form $u = u_0 \exp(ikx - i\omega_k t)$.  In the limit
of vanishing $\alpha$, we find
\begin{equation}
  \label{eq:dispersion}
  k^2 = \omega_k^2\left(1 + {\lambda \over 1 - i\omega_k\tau}\right).
\end{equation} 
The wave speed $c$ is
\begin{equation}
  \label{eq:wave_speed}
  c \equiv \lim_{k \to 0} \text{Re} \, {\omega_k \over k} = {1 \over
    \sqrt{1 + \lambda}}.
\end{equation}

It is important to recognize that, by linearizing, we have reduced our
system to a conventional model of viscoelasticity.  The solution of
the time-dependent version of (\ref{eq:lindotu}) can be written in the
familiar form
\begin{equation}
  \label{eq:epst}
  \epsilon_{tot}(t) = {1\over c^2} \sigma(t) - \lambda \int_{-\infty}^t
  dt' \exp\left[ -{1 \over \tau}(t - t')\right] \dot\sigma(t') 
\end{equation}
where $\epsilon_{tot}$ is the total (elastic plus plastic) strain, and
$\sigma = \partial u_{el}/\partial x$ is the stress in dimensionless
units.  Equivalently,
\begin{equation}
  \label{eq:sigt}
  \sigma(t) = c^2 \epsilon_{tot}(t) + \lambda \int_{-\infty}^t dt'
  \exp\left[-{1 \over \tau c^2}(t - t')\right]
  \dot\epsilon_{tot}(t').
\end{equation}
From (\ref{eq:epst}) we find that the creep compliance --- the
variation of the strain that is produced by a unit jump in the stress
--- is
\begin{equation}
  C(t) = 1 + \lambda \left(1 - e^{-t/\tau}\right).
\end{equation}
The system exhibits unit instantaneous elasticity, following which the
strain increases on the time scale $\tau$ to its final value $1 +
\lambda = 1/c^2$. Similarly, we see from (\ref{eq:sigt}) that a unit
jump in the strain produces first an instantaneous jump in the stress,
after which the stress decreases to a constant, nonzero value.

Because our equations of motion (\ref{eq:steady_motion}) and
(\ref{eq:lindotu}) are linear, we can solve the decohesion problem
analytically. (We shall need these linear solutions in order to
interpret features of the nonlinear solutions described in the next
Section.) The analysis is particularly simple if we take the limit of
weak driving springs, $\alpha \to 0$.  To do this, we must also take
the limit $u_\infty \to \infty$ in such a way as to keep $\alpha
u_\infty$ constant.  That is, we keep $G(v)$ fixed in (\ref{eq:G}).

For $x < 0$, we can set $\alpha = 0$ immediately in
(\ref{eq:steady_motion}) and look for a solution in the form:
\begin{equation}
  u_{el} = A e^{qx}, \qquad u_{pl} = B e^{qx}.
\end{equation}
That is, we look for values of $A$, $B$, and $q$ that satisfy the
homogeneous equation:
\begin{equation}
  \label{eq:matrix-neg}
  \left(
    \begin{array}{cc}
      -\beta^2 q^2 + \kappa^2, & v^2 q^2 + \kappa^2 \\
      -\lambda/v\tau, & q + 1/ v\tau
    \end{array}
  \right)
  \left(
    \begin{array}{c}
      A \\ B
    \end{array}
  \right) = 0.
\end{equation}
The solvability condition for (\ref{eq:matrix-neg}) is:
\begin{equation}
  \label{eq:q_eq}  
  q^3 v\tau \beta^2 + q^2 \beta_c^2 - q v\tau \kappa^2 - {\kappa^2
    \over c^2} = 0,
\end{equation}
where
\begin{equation}
  \beta^2 \equiv 1 - v^2;~~~~ \beta_c^2 \equiv 1 - v^2/c^2.
\end{equation}
Eq.~(\ref{eq:q_eq}) has only one positive root, say $q_1$.  The
boundary condition $u_{el}(0) + u_{pl}(0) = 1$ is therefore sufficient
to determine uniquely the solution in the region $x < 0$.  We find:
\begin{equation}
  \label{eq:AB}
  A = c^2 \left({1 + q_1 v\tau \over 1 + q_1 v\tau c^2}\right);~~~~
  B = {\lambda c^2 \over 1 + q_1 v\tau c^2}.
\end{equation}

For $x > 0$, all $\kappa$'s appearing in (\ref{eq:q_eq}) must be
replaced by $\alpha$'s.  The resulting equation has two negative
solutions for $q$ which, for small $\alpha$, are:
\begin{equation}
  \label{eq:xi>0-roots}
  q_0 = - {\alpha \over \beta_c c}; \qquad q_2 = - {\beta_c^2 \over
    v\tau \beta^2} + {\cal O}(\alpha^2).
\end{equation}
We therefore construct solutions of the form:
\begin{eqnarray}
  u_{el} & = & D_1 e^{q_0x} + D_2 e^{q_2x} + c^2 u_\infty, \\
  u_{pl} & = & \left({\lambda D_1 \over 1 + v\tau q_0}\right) e^{q_0x}
  + \left({\lambda D_2 \over 1 + v\tau q_2}\right) e^{q_2x} + \cr
  && (1 - c^2) u_\infty.
\end{eqnarray}
Here, we have included the particular solutions (simple constants)
that satisfy the boundary condition $u \to u_\infty$ at $x \to
+\infty$.  We then require that $u_{el}$ and $u_{pl}$ and their
derivatives be continuous at $x = 0$.  Calculating to first order in
$\alpha u_\infty$, we find:
\begin{eqnarray}
  \label{eq:D1_and_2}
  D_1 & = & -c^2 u_\infty \left(1 + {q_0 \over q_2} \right) + A
  \left(1 - {q_1 \over q_2} \right), \\
  D_2 & = & A \, {q_1 \over q_2} + c^2 u_\infty {q_0 \over q_2},
\end{eqnarray}
where $A$ is given in (\ref{eq:AB}).  Finally,
\begin{equation}
  \label{eq:G_linear}
  K(v) \equiv \sqrt{2 G(v)} = \alpha u_\infty = {q_1 c \over \beta_c}
  \left({\beta^2 v \tau q_1 + \beta_c^2 \over c^2 v \tau q_1 +
      1}\right).
\end{equation}

\begin{figure}[htbp]
  \begin{center}
    \centerline{\epsfxsize=3.2in \epsfbox{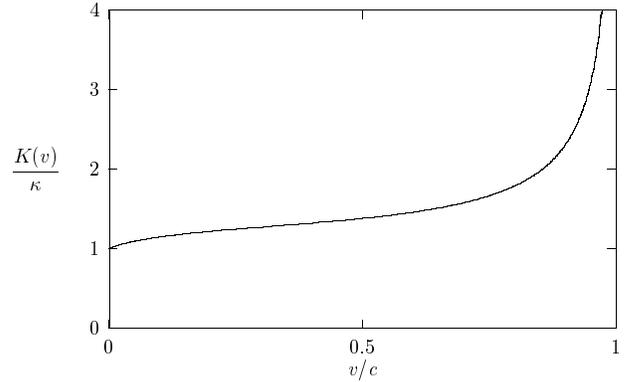}}
    \vspace{0.1in}
    \caption{Decohesion toughness $K(v)$ for the linear model with
    $\tau = 10$ and $\lambda = 1.$}
    \label{fig:K_linear}
  \end{center}
\end{figure}
We show a representative graph of $K$ as a function of the speed $v$
in Fig.~\ref{fig:K_linear}.  As expected from a model of this kind,
$K(v)$ is monotonic, diverges at $v=c$ as $1/\beta_c$, and is equal to
$\kappa$ for $v = 0$, which confirms that viscous dissipation is
negligible for slow decohesion.

\section{Solutions of the nonlinear models}
\label{sec:nonlinear}

We turn now to the nonlinear models defined by
Eqs.~(\ref{eq:steady_motion}) through (\ref{eq:steady_motion3}). These
are equivalent to a set of ordinary differential equations which, for
Model 1, are:
\begin{eqnarray}
  \label{eq:final_system}
  \beta^2 \sigma' & = & {v\over\tau}\,(\lambda\sigma - \Delta) + \kappa^2 
  \Theta(-{x})\,u + \alpha^2(u - u_\infty), \\
  \epsilon' & = & {1\over v\tau}\,(\lambda \sigma - \Delta), \\
  \label{eq:falk_Devol}  
  \Delta' & = & {1 \over v\tau}(\lambda \sigma - \Delta)[1 - \gamma_1
  \sigma \Delta], \\ 
  u' & = & \sigma + \epsilon. 
\end{eqnarray}
To obtain the analogous equations for Model 2, replace
$\gamma_1 \sigma$ in the square brackets in (\ref{eq:falk_Devol}) by
$\gamma_2 v \epsilon' = \gamma_2(\lambda \sigma - \Delta)/\tau$.

We integrate these equations numerically using a twelfth-order
predictor-corrector algorithm.  The initial conditions are $\sigma =
\epsilon = \Delta = 0$ at $x \to -\infty$, far ahead of the decohesion
front.  Our strategy is to fix the material parameters $\lambda$,
$\gamma$, and $\tau$ (or, equivalently, $c$, $\sigma_y$, and $\tau$),
the velocity $v$, and the strength of the driving springs $\alpha^2$,
and to adjust $u_\infty$ to obtain a solution with the property that
$u(0) = 1$ and $u \to u_\infty$ as $x \to +\infty$. Such a solution
always exists. Recall that, in our analogy with the crack propagating
in a prestressed strip, the parameter $u_\infty$ is analogous to the
displacement of the strip edges far ahead of the crack tip.  We
therefore might think of our procedure as adjusting the driving stress
on the strip to achieve a certain velocity of fracture propagation.

Before looking in detail at these solutions, consider the following
thought experiment.  Imagine that we start with a static, unstressed
system and $u_\infty = 0$.  Suppose also that the cohesive springs act
only for $x < 0$; that is, we arbitrarily disconnect them for $x > 0$.
Let us now increase $u_\infty$ from zero quasistatically.  In this
limit of infinitesimally slow displacement and a fixed position of the
decohesion front, the nonlinear models are indistinguishable from the
linear model as long as the stress in the membrane nowhere exceeds the
plastic yield stress.  This is because, for $\sigma < \sigma_y$, the
quasistatic system must stay arbitrarily close to the nonflowing state
with $\epsilon = \Delta = \lambda \sigma$ and the nonlinear term in
(\ref{eq:delta_motion}) is irrelevant.

The linear theory tells us that the largest stress occurs at $x = 0$
where, for this quasistatic situation, it has the value $\sigma_{max}
= \alpha u_\infty \kappa$.  At this point, the displacement of the
membrane is $u(0) = \alpha u_\infty/c$.  Clearly, the behavior of this
system depends sensitively on whether or not $\sigma_{max}$ exceeds
the plastic yield stress $\sigma_y$ before $u(0)$ reaches the breaking
point $u(0) = 1$.  Thus the critical value of the yield stress that
marks some kind of quasistatic boundary between brittle and ductile
behavior of these models is $\sigma_y = \kappa c$.

If $\sigma_y > \kappa c$, then the threshold for propagating
decohesion is reached before any plastic flow occurs, and we deduce
that both nonlinear models behave much like the linear model for small
enough speeds $v$ --- \ie\ they are ``brittle.''  On the other hand,
if $\sigma_y < \kappa c$, then plastic flow occurs before the leading
cohesive spring breaks.  In this case, the two nonlinear models
behave differently from one another.

In Model 1, plastic flow must begin as soon as the maximum stress
reaches the yield stress.  As shown in Fig.~\ref{fig:mod1_flow}, the
flowing and nonflowing states cross at this point.  If $u_\infty$ is
increased very slowly beyond this point, just as in conventional
models of plasticity, the stress at $x = 0$ remains fixed at
$\sigma_y$.  The displacement $u(0)$ also remains fixed at its value
below the decohesion threshold, $u(0) = 1$, because no additional
stress can be applied to stretch the cohesive springs. The only thing
that can happen is that, as $u_{\infty}$ continues to grow, the
material in the region $x > 0$ deforms plastically.  Thus, decohesion
is not initiated, but an indefinitely large amount of plastic work is
done on the system.  We therefore anticipate that the decohesion
toughness $K(v)$ for Model 1 must diverge at $v = 0$ whenever
$\sigma_y < \kappa c$.

The most interesting questions, of course, have to do with the
behavior at nonzero propagation speeds $v$, where the quasistatic
approximations are not necessarily valid.  In general, our decohesion
criterion $u(0) = 1$ implies that the breaking stress $\sigma(0)$
increases with increasing $v$. (We shall not consider a stress-based
criterion, which might be simpler in some respects.)  Actually, at
nonzero speeds, the stress reaches its maximum some distance behind
the decohesion front.  In the linear version of the model, this
maximum stress diverges at $v = c.$ Thus, even if the maximum stress
$\kappa c$ is less than $\sigma_y$ at $v = 0$, it will become greater
than $\sigma_y$ at some onset speed for plastic flow that we shall
call $v_p$.  At speeds larger than $v_p$, the system must deform
plastically.

\begin{figure}[htbp]
  \begin{center}
    \centerline{\epsfxsize=3.2in \epsfbox{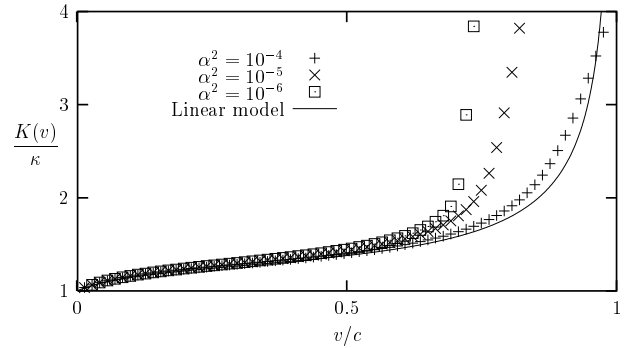}}
    \vspace{0.1in}
    \caption{Decohesion toughness $K(v)$ with $\sigma_c = 3 \kappa c$
      and $\tau = 10,$ for three values of the driving $\alpha$.  The
      limit of weak driving $\alpha \to 0$ exists only for $v < v_p
      \approx 0.73c$.}
    \label{fig:Kmid_new}
  \end{center}
\end{figure}
\begin{figure}[htbp]
  \begin{center}
    \centerline{\epsfxsize=3.2in \epsfbox{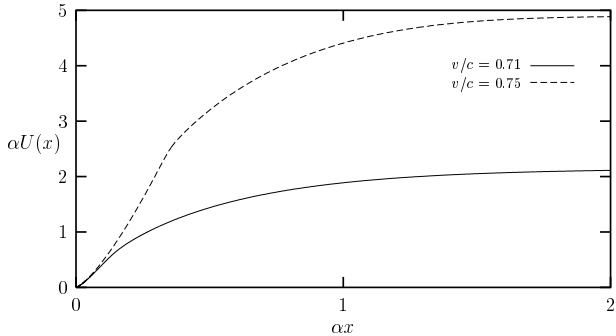}}
    \vspace{0.1in}
    \caption{Total displacement $U(x)$ as a function of $x,$ both
      measured in units of $\alpha^{-1},$ for speeds just below and
      just above $v_p.$ A considerable increase in the driving force
      $\alpha U_\infty$ is needed to increase the decohesion speed by
      4\%.}
    \label{fig:disp1}
  \end{center}
\end{figure}
\begin{figure}[htbp]
  \begin{center}
    \centerline{\epsfxsize=3.2in \epsfbox{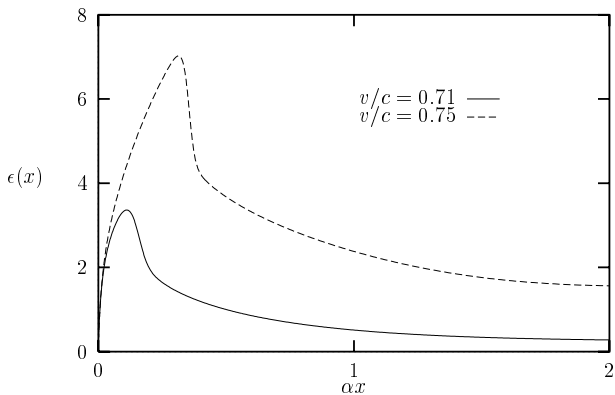}}
    \vspace{0.1in}
    \caption{Plastic strain $\epsilon(x)$ as a function of $\alpha x$
      for speeds just above and just below $v_p.$ Note that the peak
      in $\epsilon(x)$ for $v/c = 0.75 > v_p$ is more than twice that
      of the peak for $v/c = 0.71 < v_p,$ and the level of plastic
      strain far behind the decohesion front changes by a factor of
      about 5.}
    \label{fig:eps1}
  \end{center}
\end{figure}

To see what happens at $v_p$, we must look at the numerical solutions
of Equations (\ref{eq:final_system}) through (\ref{eq:falk_Devol}).
We continue to consider only Model 1 for the moment.  In
Fig.~\ref{fig:Kmid_new}, we show the decohesion toughness $K(v) =
\alpha u_\infty$ as a function of $v/c$, for $\sigma_y = 3\kappa c$,
$\tau = 10$, $\lambda = 1$, and for three different values of $\alpha$.
For comparison, we also show $K(v)$ for the linear theory.  The most
striking feature is that these four curves are almost coincident for
$v/c$ less than a critical value of about $0.73$; but they break away
from the linear theory at larger speeds, the systems with smaller
$\alpha$ being the most dissipative.  We see even more explicitly what
is happening in Figs.~(\ref{fig:disp1}) and (\ref{fig:eps1}) where we
have plotted the total displacement $u(x)$ and the plastic strain
$\epsilon(x)$ for $\alpha = 10^{-3}$, and for two velocities $v/c =
0.71$ and $0.75$, just below and just above the critical speed
respectively.  Note that the plastic strain is very much larger for
the slightly larger value of $v$.  In that case, the plastic strain
grows almost linearly with $x$ before it reaches its peak.  The
spatial extent of the region in which this plastic strain accumulation
happens seems to scale linearly with $\alpha^{-1}$.  Note also that
decohesion produces a residual plastic deformation and, accordingly, a
residual stress (not shown here) that persist infinitely far behind
the front.

\begin{figure}[htbp]
  \begin{center}
    \centerline{\epsfxsize=3.2in \epsfbox{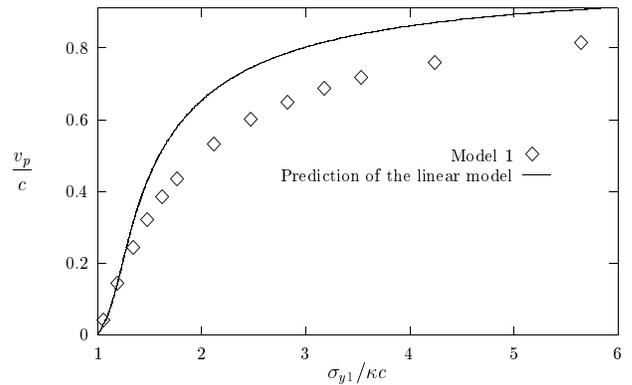}}
    \vspace{0.1in}
    \caption{Onset velocity $v_p$ as a function of the yield stress
      for Model 1.}
    \label{fig:v_p1}
  \end{center}
\end{figure}

We deduce from this data that the $\alpha \to 0$ limit exists only for
$v < v_p$.  That is, if we try to drive decohesion at a speed greater
than $v_p$ with springs of arbitrarily small force-constant
$\alpha^2$, the dissipated energy per unit advance of the front grows
without bound.  For Model 1, $v_p$ is the speed at which the maximum
stress just exceeds the plastic yield stress.  To confirm this
interpretation, in Fig.~\ref{fig:v_p1} we plot $v_p$ as a function of
$\sigma_y$ and compare this with the prediction of the linear model
for the maximum stress
\begin{equation}
  \sigma_{max} = {c \over \beta_c} \alpha u_\infty \approx
  \sigma_y~~~\text{at}~~~v = v_p(\sigma_y),
\end{equation}
where $\alpha u_\infty$, is given in Eq.~(\ref{eq:G_linear}).
Agreement with the linear theory becomes exact in the limit of slow
decohesion.  We can also qualitatively understand the fact that the
prediction of the linear model consistently overestimates the onset
velocity $v_p$.  When the nonlinear term can be treated
perturbatively, it leads to a decrease in the relaxation rate of
$\Delta$, since the right hand side of Eq.~(\ref{eq:falk_Devol}) is
reduced.  The right hand side of the equation for the derivative of
the stress (\ref{eq:final_system}) is therefore increased thus
allowing the stress to reach a higher level before the $\alpha^2 (u -
u_\infty)$ term reverses the sign of $\sigma'$.  As a result, the
maximum stress in a nonlinear system reaches yield at a lower
velocity than in the corresponding linear system.

So far, we have looked in detail only at the physically less realistic
situation in which the plastic yield stress is higher than the
breaking stress at the $v = 0$ threshold for decohesion.  We now
consider the case where the yield stress is lower than the breaking
stress.  In Model 1, if $\sigma_y < \kappa c$, then we are always in
the regime where the $\alpha \to 0$ limit fails to exist.  In
examining the behavior in this regime, therefore, we choose a small,
fixed value for $\alpha$, specifically $\alpha = 0.01$, and look at
various values of other parameters.

\begin{figure}[htbp]
  \begin{center}
    \centerline{\epsfxsize=3.2in \epsfbox{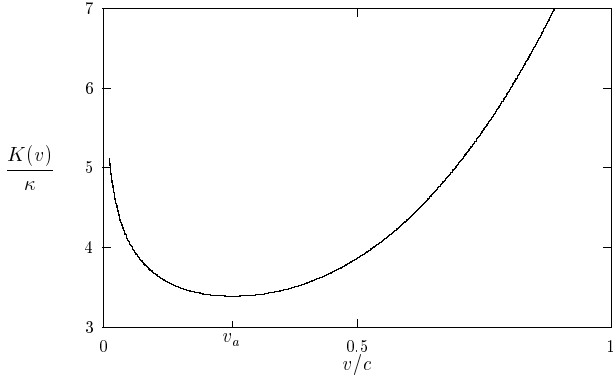}}
    \vspace{0.1in}
    \caption{Decohesion toughness $K(v)$ for Model 1 in the case
      $\sigma_{y1} = 0.5 \kappa c$ for $\alpha = 0.01.$}
    \label{fig:mod1_K}
  \end{center}
\end{figure}
\begin{figure}[htbp]
  \begin{center}
    \centerline{\epsfxsize=3.2in \epsfbox{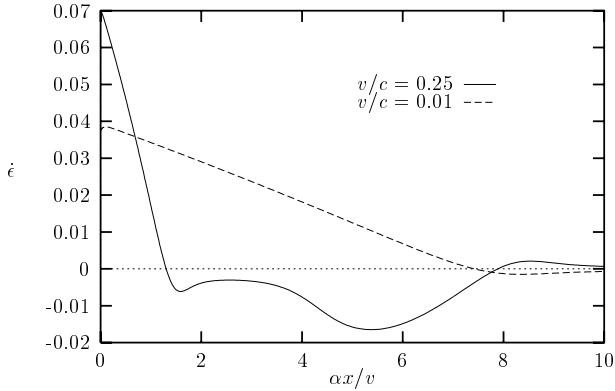}}
    \vspace{0.1in}
    \caption{Plastic strain rate as function of time $x/v$ after the
      passage of the decohesion front in Model 1.  The parameters are
      the same as in Figure \ref{fig:mod1_K}.  Plastic flow persists
      much longer for slow decohesion.  Also note that plastic strain
      recovery is appreciable only when decohesion is fast.}
    \label{fig:mod1_str_rate}
  \end{center}
\end{figure}

Fig.~\ref{fig:mod1_K} is a graph of $K(v)$ as a function of $v/c$
for $\sigma_y = 0.5 \kappa c$.  The most interesting feature of this
graph is that $dK/dv < 0$ for speeds $v$ between zero and, say, $v_a$.
Propagation at speeds in that interval must be unstable; thus, if we
increase the driving parameter $u_\infty$ to some value such that $K =
\alpha u_{\infty} > K(v_a)$, then $v$ must jump to some value on the
high-speed, stable branch of this curve.  Conversely, if we decrease
the driving force so that $\alpha u_\infty < K(v_a)$, then decohesion
must stop abruptly.  As anticipated, the decohesion toughness is large
at small speeds because the plastic strain relaxes very slowly near
threshold, and the flowing region extends far behind the decohesion
front.  At larger speeds, the deformation is more localized, and there
is less dissipation.  At yet larger speeds, of course, the driving
force must increase in order to make the front move at speeds
comparable to $c$.  To illustrate these variations in the plastic
deformation explicitly, in Fig.~\ref{fig:mod1_str_rate} we plot the
plastic strain rate $\dot \epsilon$ as a function of $x/v$.  This
figure can be interpreted as the plastic strain rate as a function of
time after passage of the decohesion front.

We turn now to the properties of Model 2, which must behave in a less
conventional manner according to Fig.~\ref{fig:mod2_flow}.  Even
when the breaking stress exceeds the plastic yield stress, $\sigma_y <
\kappa c$, the system can remain in the stationary state with $\Delta
= \Delta_{2A}$, $\dot \epsilon = 0$ so long as the stress is raised
sufficiently slowly.  No plastic flow occurs, and the decohesion
toughness at $v \to 0$ must be $\kappa$, just as in the linear model.

To predict the onset of plastic flow in Model 2, \ie\ to compute the
analog of $v_p(\sigma_y)$, we can use the linear theory to estimate
when the nonlinear term in the equation of motion
(\ref{eq:falk_Devol}) becomes non-negligible.  It is useful to carry
out this exercise for both models. For Model 1, validity of the linear
approximation requires
\begin{equation}
  \label{eq:ineq1}
  \gamma_1 \sigma \Delta \cong \gamma_1 u'_{el} u'_{pl} \ll 1.
\end{equation}
Our linear analysis tells us that
\begin{equation}
  \gamma_1 u'_{el} u'_{pl} \approx \gamma_1 \lambda (\alpha c
  u_\infty)^2 \left(1 - {v\tau q_1 \over 1 + v\tau q_1}
    e^{-x/v\tau}\right).
\end{equation}
This quantity vanishes at the decohesion front, $x = 0$, and rises
monotonically to a constant as $x \to \infty$.  We know that $\gamma_1
\lambda = 1/\sigma_y^2$ and, for $\kappa v \tau \ll 1$, $\alpha
u_\infty \approx \kappa$.  Thus the inequality in (\ref{eq:ineq1})
reduces to $\sigma_{y1} \gg \kappa c$, consistent with our quasistatic
analysis for this model.

Model 2 behaves differently.  The analog of the inequality
(\ref{eq:ineq1}) is
\begin{equation}
  \label{eq:ineq2}
  \gamma_2 v \epsilon' \Delta \cong \gamma_2 v u_{pl}' u_{pl}'' \ll 1.
\end{equation}
From the linear analysis, we find
\begin{eqnarray}
  \gamma_2 v u_{pl}' u_{pl}'' & \approx & \gamma_2 v \lambda^2 q_1^3
  c^4 {(1 + v\tau q_1) \over (1 + v\tau q_1 c^2)^2} e^{-x/v\tau}
  \times \cr && \left(1 - {v\tau q_1 \over 1 + v\tau q_1}
    e^{-x/v\tau}\right).
\end{eqnarray} 
Now the nonlinear correction is localized in a finite region whose
size is of order $v\tau$ near the decohesion front.  We use $\gamma_2
\lambda^2 = 4\tau/\sigma_{y2}^2$.  Then, in the case $\kappa v\tau \ll
1$, the inequality (\ref{eq:ineq2}) becomes
\begin{equation}
  \label{eq:ineq2b}
  v \tau \ll {\sigma_{y2}^2 \over 4 c \kappa^3},
\end{equation}
or, equivalently,
\begin{equation}
  \label{eq:ineq2a}
  {\kappa v\tau \over c} \ll \left[{\sigma_{y2} \over 2
      \sigma(0)}\right]^2.
\end{equation}
Both sides of (\ref{eq:ineq2a}) are accurate only for $\kappa v\tau
\ll 1$.  In the opposite limit, $\kappa v\tau \gg 1$, (\ref{eq:ineq2})
reduces simply to $\sigma_{y2} \gg \kappa$. The important point is that,
when $v$ is sufficiently small in Model 2, brittle behavior can occur
for values of the plastic yield stress $\sigma_{y2}$ that are smaller
than the decohesion stress $\sigma(0)$.  The right-hand side of
(\ref{eq:ineq2b}) gives us an upper bound for $v_p \tau$ for Model 2.

\begin{figure}[htbp]
  \begin{center}
    \centerline{\epsfxsize=3.2in \epsfbox{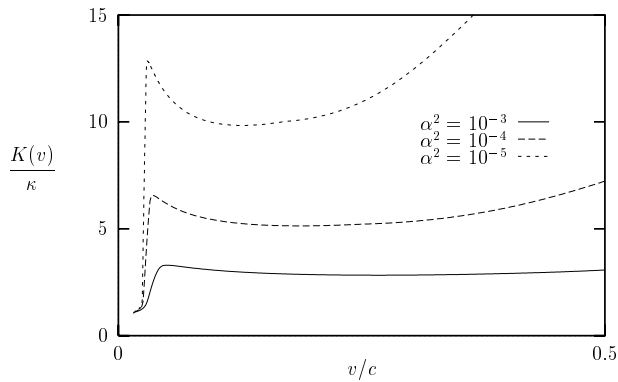}}
    \vspace{0.1in}
    \caption{Decohesion toughness $K(v)$ in Model 2 for three
      different values of $\alpha.$ Other parameters are $\sigma_{y2}
      = 0.8 \kappa c$, $\lambda = 1$ and $\tau = 4$.}
    \label{fig:mod2_K}
  \end{center}
\end{figure}
\begin{figure}[htbp]
  \begin{center}
    \centerline{\epsfxsize=3.2in \epsfbox{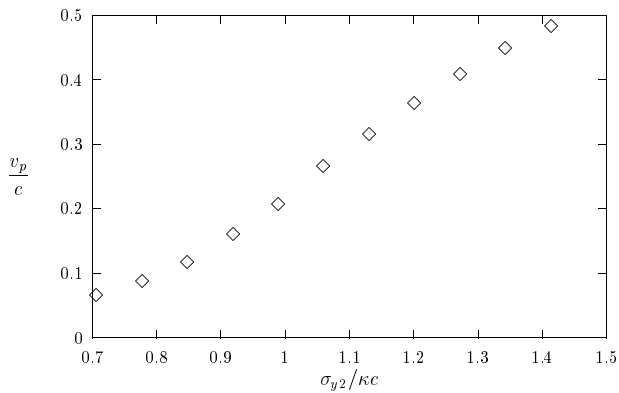}}
    \vspace{0.1in}
    \caption{Onset velocity $v_p$ in Model 2 for $\lambda = \tau = 1$
      as a function of the yield stress $\sigma_{y2}.$ Lower yield
      stresses are numerically difficult to access.}
    \label{fig:mod2_v_p}
  \end{center}
\end{figure}

These features of the behavior of Model 2 are confirmed by our
numerical results.  In Fig.~\ref{fig:mod2_K}, we show the decohesion
toughness $K(v)$ as a function of $v$ for $\sigma_{y2} = 0.8 \kappa c$
for three different values of $\alpha^2$, $\lambda = 1$ and $\tau =
4$.  As expected, $K(0) = \kappa$.  There is a stable region at small
$v$ where $dK/dv > 0$ and in which the limit $\alpha \to 0$ exists.
The the onset of plastic flow speed, and the failure of that limit,
occurs at $v = v_p$ where $K(v)$ first begins to rise sharply.  Beyond
its maximum, the fracture toughness in Model 2 behaves qualitatively
like Model 1.  That is, for small but nonzero values of $\alpha$,
there is an unstable region in which $K(v)$ decreases for increasing
$v$.  At yet larger values of $v$, $K(v)$ rises again and diverges as
$v$ approaches $c$ in the limit of $\alpha \to 0$.  As shown in
Fig.~\ref{fig:mod2_v_p}, the plastic flow onset velocity $v_p$ in
Model 2 vanishes only as the yield stress $\sigma_y$ is reduced to
zero.  Its behavior for small yield stresses is consistent with the
prediction of the perturbation theory Eq.~(\ref{eq:ineq2b}).

\section{Discussion}
\label{sec:discussion}

We have explored these one-dimensional models of decohesion primarily
as an attempt to understand how a fully dynamic description of
viscoplasticity might play a role in theories of dynamic fracture.  We
are particularly interested in how the concepts of brittleness and
ductility will emerge in such theories.

To begin this discussion, consider one way in which we might expect
the brittle-ductile transition to appear in a theory of, say, mode I
fracture along the centerline of an infinitely long, two-dimensional
strip.  Let the width of the strip be $2 W$, and suppose that the
driving force is produced by a rigid displacement of size $u_\infty$
at its edges. In the case of brittle fracture, the stress-intensity
factor at the tip of the crack is proportional to $K_I =
u_\infty/\sqrt W$. Infinitely far behind the crack tip, the
steady-state crack opening displacement is $2 u_\infty$. In the limit
$W \to \infty$, $K_I$ remains fixed for a fixed speed of crack
propagation, and thus the ratio $u_\infty/W$ vanishes.  The crack
remains sharp and narrow on the macroscopic scale $W$.

The extreme ductile version of this situation is one in which the
solid is replaced by a viscous fluid, and the crack becomes a finger
in a Hele-Shaw cell. In this case, the steady-state finger has a width
$u_\infty$ of approximately $W$, and the dissipation rate,
proportional to $K_I^2$, diverges as $W \to \infty$.  In any real two
or three dimensional solid, of course, the plastic yield stress is
nonzero.  Therefore, as we increase $W$ at fixed crack speed --- no
matter whether the crack is advancing in a brittle or ductile manner
--- we must eventually get to the point where the stresses far away
from the crack tip drop below the plastic yield stress.  The
dissipation may become very large, but it remains finite when
$W \to \infty$.

In our one-dimensional model, the closest available analog of the
width $W$ is the length scale $\alpha^{-1}$.  However, we have no
analog of the stress concentration that is produced by a real second
dimension, and thus we have no way in which the far-field stresses can
be made arbitrarily small --- less than the plastic yield stress ---
by taking the limit of a large system.  In its brittle mode, as we
have seen, the analog of the stress-intensity factor for our
decohering membrane is $K(v)= \alpha u_\infty \approx u_\infty/W$;
this quantity remains finite at fixed $v$ in the limit $\alpha \to 0$.
In its ductile mode, however, our system is behaving more like a
finger in a Hele-Shaw cell than a crack in a solid strip.  As soon as
plastic flow starts at any point, the dissipation rate diverges as
$\alpha \to 0$.  In short, the distinction between brittle and ductile
failure in this system must be qualitatively unlike that which occurs
in real fracture.  To understand the latter, we shall have to carry
out fully two-dimensional investigations.

What, then, are the lessons to be learned from this exercise?  What
questions does it raise?  We have confirmed, as expected, that the FL
model of viscoplasticity produces both brittle and ductile propagating
failure modes.  In this one-dimensional version, the transition
between brittle and ductile behavior is perfectly sharp and well
defined; it is distinguished by the divergence of the decohesion
toughness in the limit $\alpha \to 0$.  Is there any such sharp
distinction in higher dimensions?  The behavior of our one-dimensional
models, especially at small propagation speeds $v$, is strongly
sensitive to our choice of the nonlinear coupling between the plastic
strain rate and the new variable $\Delta$ that describes the internal
state of the system.  It will be important to learn whether more
realistic models in higher dimensions exhibit similar sensitivity to
details of the mechanisms that control plastic flow.

Perhaps the most interesting but problematic aspect of our results is
related to that sensitivity.  Both of the models discussed here
exhibit unstable steady-state solutions at low propagation speeds.  We
know that these solutions are unstable because the decohesion
toughness $K(v)$ decreases with increasing $v$.  In both models,
$K(v)$ rises again at higher speeds, and the stable high-$v$ solutions
are ductile.  In Model 2, however, there is also a stable small-$v$
solution that is brittle.  That is, there exists a range of values of
$K(v)$ for which there are two stable steady-state solutions, one
brittle and one ductile.  Within such a range of driving forces, the
system is likely to exhibit complex, non-steady behavior.  What might
be the analog of such behavior in two-dimensional models of fracture?
Might there be situations in which multiple solutions exist but both
are brittle?  Or might the slow solution be the ductile one?  More
generally, might the new dynamics emerging from the FL model of
viscoplasticity be a clue for understanding the complex instabilities
and different modes of behavior observed in real fracture?

\section*{Acknowledgments}
\label{sec:acknow}

We thank M.~L.~Falk for many stimulating conversations.  This research
was supported by DOE Grant No. DE-FG03-84ER45108 and also in part by
NSF Grant No. PHY94-07194.

\bibliographystyle{prsty}
\bibliography{plasticity}

\end{document}